\documentclass[a4paper,12pt]{article}  
\usepackage{graphicx,sectsty,longtable,tocloft,color}
\usepackage{epstopdf}
\usepackage{sidecap}
\usepackage{subfig}
\usepackage{xspace}
\usepackage{mydefs}
\usepackage{cite}
\usepackage{lineno}
\usepackage{multirow}
\usepackage[left=2cm,right=2cm,top=2cm,bottom=2cm]{geometry}
\title{Boosted Higgs $\rightarrow b\bar{b}$ in vector-boson associated production at 14 TeV}

\author{Jonathan M. Butterworth, In\^{e}s Ochoa, Tim Scanlon \\
Department of Physics and Astronomy, University College London, \\
Gower St., London, WC1E 6BT, UK
}

\begin{document}

\maketitle

\begin{abstract}
The production of the Standard Model Higgs boson in association with a vector boson, followed by the dominant decay to $H \rightarrow b\bar{b}$, is a strong prospect for confirming and measuring the coupling to $b$-quarks in $pp$ at $\sqrt{s}=14$~TeV. We present an updated study of the prospects for this analysis, focussing on the most sensitive highly Lorentz-boosted region. The evolution of the efficiency and composition of the signal and main background processes as a function of the transverse momentum of the vector boson are studied covering the region $200-1000$~GeV, comparing both a conventional dijet and jet substructure selection. The lower transverse momentum region ($200-400$~GeV) is identified as the most sensitive region for the Standard Model search, with higher transverse momentum regions not improving the statistical sensitivity. For much of the studied region ($200-600$~GeV), a conventional dijet selection performs as well as the substructure approach, while for the highest transverse momentum regions ($> 600$~GeV), which are particularly interesting for Beyond the Standard Model and high luminosity measurements, the jet substructure techniques are essential.
\end{abstract}

\section{Introduction}
\label{sec:intro}

Following the discovery of a Higgs boson~\cite{Aad:2012tfa,Chatrchyan:2012ufa} with a mass of around 125~GeV principally via its decay to gauge 
bosons ($\gamma, Z, W$), the task of confirming and then measuring the presumed-dominant decay to $b\bar{b}$ remains a priority and a challenge. 
The most sensitive searches for this decay mode to date are in the ``boosted'' 
region of the $VH$ production channel - that is, when the Higgs ($H$) and the vector boson ($V$) both have transverse momentum $p_T > 200$~GeV or so. 
Two approaches can be used to reconstruct the Higgs boson in this region: two nearby, separate ``resolved'' $b$-jets can be identified, 
or a single ``fat'' jet can be found and decomposed using jet substructure techniques.

The use of jet substructure techniques to identify hadronically-decaying boosted, massive particles was suggested some time before 
the start-up of the Large Hadron Collider~\cite{Seymour:1993mx,Butterworth:2002tt}, and has seen much phenomenological and 
experimental activity and progress over recent years (see \cite{Altheimer:2013yza} for a recent overview). 
Jet substructure and/or ``grooming'' techniques have claimed many successes in recent measurements and searches, and in particular 
have been shown to not only be robust against soft QCD effects such as underlying event and multiple proton-proton interactions (pile-up), 
but in some cases an essential tool for reducing their impact~\cite{Aad:2013gja,Chatrchyan:2013vbb}. 

An early expectation was that boost, and hence jet substructure, would be important for identifying the $b\bar{b}$ decay mode of a low-mass 
Higgs boson~\cite{ref:boostedhiggsprl}. The searches to date for this decay mode using LHC data~\cite{Chatrchyan:2013zna,Aad:2014xzb}
indeed gain most of their sensitivity from the boosted region - in which the Higgs and the vector boson both 
have transverse momentum $p_T > 200$~GeV - but do not exploit jet substructure. 
One reason for this is the excellent performance of the \hbox{anti-${k_T}$} jet algorithm~\cite{ref:antikt} used by both ATLAS and CMS. 
When run with a radius parameter of $R = 0.4$ (ATLAS) or $0.5$ (CMS), a good mass resolution is obtained 
along with well-defined jet separation, even for jet pairs which are quite boosted.
Another is the fact the mass of the Higgs boson, at 125~GeV, turned out to be
towards the high end of the applicability of the jet substructure methods, which would have been most effective for a 115~GeV Higgs boson. 
Finally, a major reason is assumed to be the fact that the LHC has not yet 
reached its design energy of 14~TeV, but ran in 2010 and 2011 at centre-of-mass energies of 7~TeV, and in 2012 at 8~TeV. 
The lower centre-of-mass energy shifts the balance in favour of the 
un-boosted region of phase-space with respect to the expectations at 14~TeV, reducing the high-$p_T$ fraction of the cross section substantially. 

We examine these assumptions, and re-evaluate the potential impact of using jet substructure techniques to 
decompose a large-radius ``fat'' jet on the search for the $H \rightarrow b\bar{b}$ decay in the $VH$ channel in the 14~TeV era, 
by conducting a particle-level study of boosted $WH,H\rightarrow b\bar{b}$ production. Although we only consider the $WH,H\rightarrow b\bar{b}$ channel, we expect the conclusions on the resolved and jet substructure approaches, to be largely applicable to the $ZH,H\rightarrow b\bar{b}$ channels.

\section{Event Generation and Selection}
\label{sec:sel}
Candidate $W$ bosons are identified by requiring a muon with $p_{T}$$>20$~GeV and absolute pseudorapidity $|\eta|<3.0$, 
as well as a neutrino with $p_{T}$$ > 20$~GeV. 
Only events in which the $p_T$ of the $W$ is greater than 200~GeV are considered.  It is assumed, based on previous measurements, that the presence of a high $p_{T}$ lepton, as well as two highly boosted $b$-jets, allows for very efficient triggering, and that there is negligible efficiency loss due to the trigger within the acceptance.

Two jet algorithms are used in this study: \hbox{anti-${k_T}$} $R=0.4$ and 
Cambridge/Aachen~\cite{ref:cambridgeaachen} $R=1.2$ split and filtered~\cite{ref:boostedhiggsprl} jets. The analysis was performed using a Rivet~\cite{Buckley:2010ar} routine, making extensive use of fastjet~\cite{Cacciari:2011ma}\footnote{The Rivet analysis code 
is available from the authors on request.}.
 
The geometrical matching of jets or subjets to $B$-hadrons is performed by requiring a $\Delta R$ 
condition\footnote{Defined as $\Delta R = \sqrt{(\Delta\phi)^{2}+(\Delta\eta)^{2}}$, where $\phi$ is the azimuthal angle.} on their overlap, chosen to be less than 0.4 or 0.3, respectively. 
A variable-$R$ matching was also tried for the subjets, where $R$ was defined as the subjet radius, but was found to bring no significant improvement to the analysis sensitivity. This statement is in part dependent on the background composition, and in particular if charm rejection were to be significantly improved, variable-$R$ matching could bring benefits since it rejects more genuine $b\bar{b}$ events.
If more than one $B$-hadron overlaps, the closest is chosen, and the matching continues with the remaining 
hadrons. Only $B$-hadrons with \mbox{$p_{T}$$ > 5$ GeV} are considered. 
If a jet or subjet is not matched to a $B$-hadron, an additional check is performed with charm hadrons, 
to allow the experimental charm-quark mis-tag rate to be estimated.
If both matching conditions fail, the jet is labelled as `light'.

Higgs boson candidates are selected in two different ways.
In the resolved approach, the following requirements are applied:
\begin{itemize}
\item At least two \hbox{anti-${k_T}$} $R=0.4$ jets with $p_{T}$$> 20$ GeV, $|\eta|<3$;
\item $\Delta R < 1.4$ between the two leading \hbox{anti-${k_T}$} jets;
\item Each of the two leading jets is matched to a $B$-hadron.
\end{itemize}
In the substructure approach the following requirements are applied:
\begin{itemize}
\item At least one Cambridge/Aachen split and filtered jet with $p_{T}$$ > 180$ GeV, $|\eta|<3$;
\item The two subjets with highest $p_{T}$ in the leading Cambridge/Aachen split and filtered jet are each matched to a $B$-hadron.
\end{itemize}

After this event selection the dominant backgrounds are top-pair production ($t\bar{t}$) and $W+b\bar{b}$, 
with additional contributions from $Wt$ and $WZ$ processes. 
In addition to the vector boson candidate selection, a veto on the number of jets in the event 
is applied to suppress these backgrounds, such that events with more than three \hbox{anti-${k_T}$} jets with $p_{T}$$ > 20$ GeV and $|\eta| < 5$ 
are rejected, and the sub-subleading \hbox{anti-${k_T}$} jet, if present, is required to be in the forward region ($|\eta| > 3.0$) or to have 
low transverse momentum (less than 10\% of $p_{T}(W)$). These cuts are used to make a more realistic 
estimate of the signal-to-background ratio and significance.
They carry significant theoretical uncertainties and experimental challenges, but do not strongly affect the comparison between the resolved and substructure approaches since 
they are the same for both\footnote{Also, we note that they would not be as important in the $ZH$ channel ($Z\rightarrow l^+l^-$), since the top background is suppressed.}. 

In the simulation of signal and backgrounds, the calculation of the matrix elements is performed with a\textsc{mc@nlo}\cite{Alwall:2014hca}, 
including NLO corrections in QCD. 
The description of the processes is improved by matching the NLO calculation with a parton-shower program, in this case \textsc{herwig++}\cite{Frixione:2002ik,Frixione:2010ra,Bahr:2008pv}, which also includes models of the underlying event and hadronisation. 
The renormalisation and factorisation scales are dynamically defined as the sum of the transverse masses of all final state particles and 
partons\footnote{The other parameters used are: $M_Z = 91.19$~GeV; $G_{F} = 1.166\times10^{-5}$~GeV$^{-2}$; $\alpha_{S}(M_{Z}) = 0.118$; $\alpha_{EW}(M_{Z}) = 1/132.5$; $M_H = 125$~GeV; $m_b = 4.70$~GeV; $m_t = 174.30$~GeV.}. 
For all processes except $Wt$, the decays of the $t, W$ and $H$ are simulated using MadSpin~\cite{Artoisenet:2012st}, 
considering the $t\rightarrow Wb$, $W\rightarrow\mu\nu$, $W\rightarrow q\bar{q}'$ and $H\rightarrow b\bar{b}$ decay modes, with the branching ratios set to 1.0, 0.11, 0.68 and 0.58, 
respectively. For the $Wt$-channel event generation, the interference with $t\bar{t}$ is dealt with by the Diagram Removal scheme \cite{Frixione:2008yi}, and the $W$ and $t$ decays are performed by \textsc{herwig++}.
Multi-jet processes can also be a background to $WH,H\rightarrow b\bar{b}$ searches. 
However, their contribution is negligible in the boosted region and is disregarded here. 

Pile-up is not simulated. Studies using full detector simulation indicate that jet grooming techniques can remove effects of pile-up to a large extent~\cite{Altheimer:2013yza}, even under extreme conditions~\cite{CMS-PAS-JME-14-001}, as can pile-up subtraction techniques in the case of \hbox{anti-${k_T}$} jets~\cite{ATLAS-CONF-2014-018}. The presence of pile-up jets could also lead to a degradation in the efficiency of the jet veto cut. In this study we assume that sufficiently robust and efficient algorithms are available to reduce any efficiency loss due to pile-up jets to a negligible level. However, any efficiency loss due to pile-up jets would impact both signal and background equally, leading to a lower sensitivity overall and this would not alter the main conclusions of the study on the relative performance of the resolved and substructure approaches. 

The total rate of $t\bar{t}$ events is scaled by a factor of 1.25 based on an estimate of the impact 
of NNLO QCD contributions \cite{Czakon:2013goa}.
This assumes a uniform enhancement of the cross-section as a 
function of the top-quark $p_{T}$, 
and is therefore a conservative estimate of the 
expected behaviour\footnote{ATLAS and CMS preliminary measurements have found the $p_{T}$ spectrum of the top-quark to be softer than that 
predicted by several simulation programs~\cite{Aad:2014zka,Chatrchyan:2012saa}.}. We note that the transverse 
momentum spectrum of $VH$ production is known to be subject to significant higher order corrections~\cite{Ferrera:2011bk,Ferrera:2014lca}.

Events are weighted to take into account a $b$-tagging efficiency assumed to be 75\%, and mis-tag rates of 15\% for charm ($c$) and 1\% for all
other quarks and gluons ($l$). Although the requirement of two $b$-tagged jets reduces 
most of the $W$+jets background to $W+b\bar{b}$ events, the contribution from $W+c\bar{c}$ events is not negligible. Based on 
the yields obtained in the ATLAS result of \cite{ATLAS-CONF-2013-079}, the $W+b\bar{b}$ process is scaled by a factor of 1.2 to 
account, approximately, for the $W+c\bar{c}$ contamination. Given that in the boosted region $W+ll$ was found to only make up $\sim1\%$ of the total background, it was deemed negligible and not included in this study.  

\section{Signal Acceptance}
\label{sec:signal}

\begin{figure}
\centering
\subfloat[\label{fig:effSig}]{
	\includegraphics[width=0.48\textwidth]{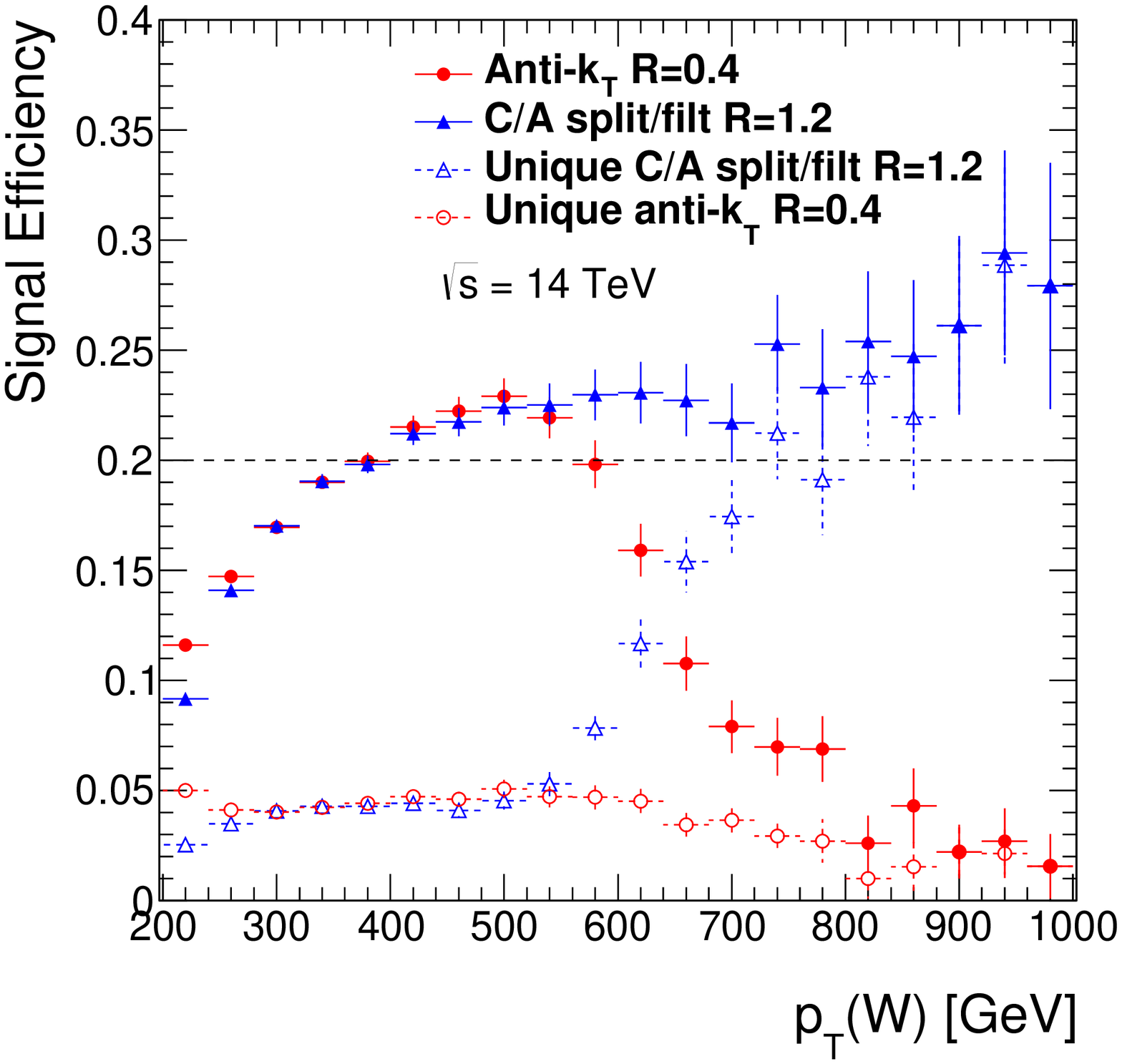}
}
\subfloat[\label{fig:xsecSig}]{
	\includegraphics[width=0.48\textwidth]{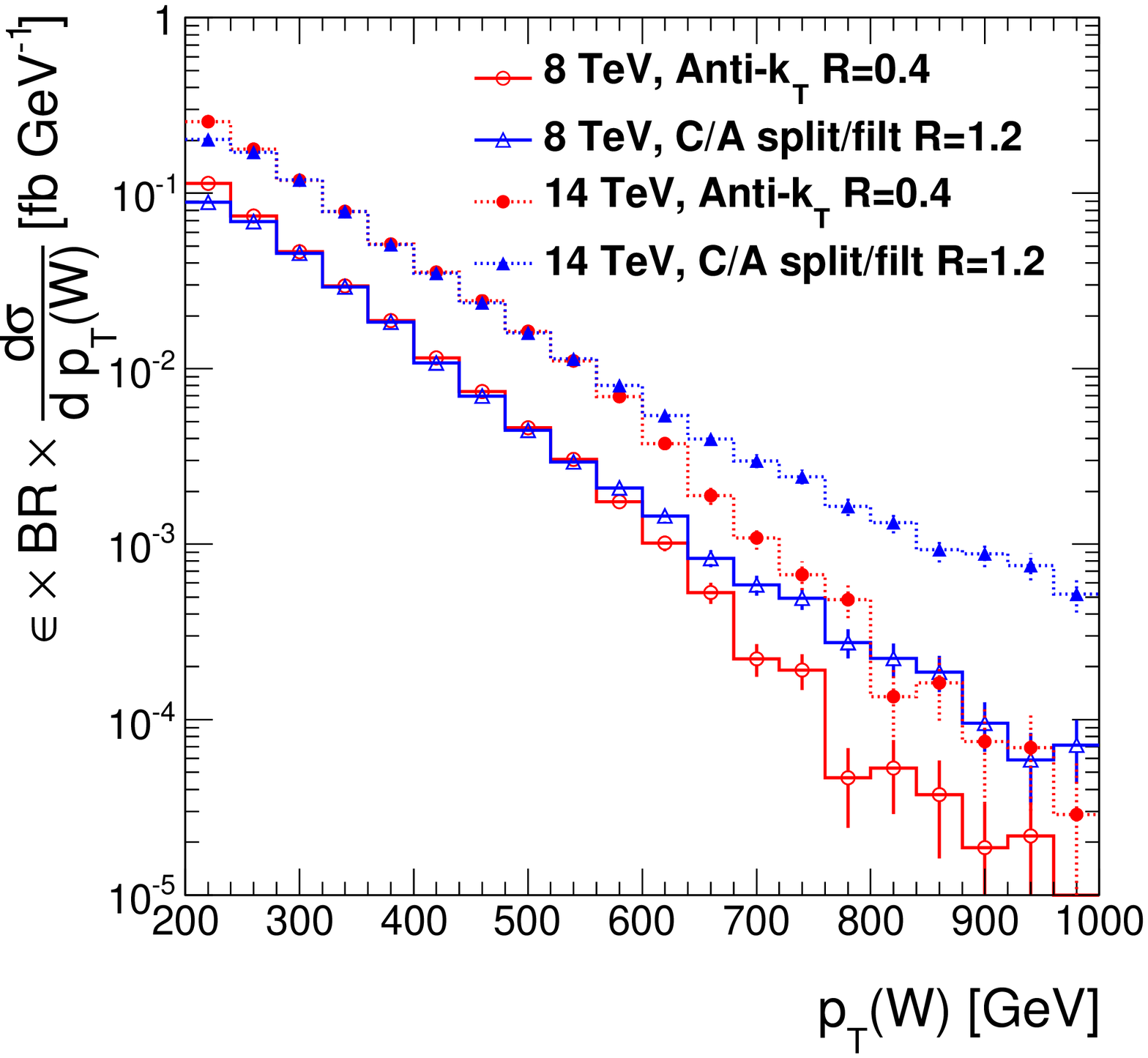}
}
\caption{\protect\subref{fig:effSig} Signal efficiency for the resolved (circles) and substructure (triangles) selections, including the fraction of signal events selected in total (solid) and uniquely (hollow) by each approach. The efficiency is evaluated with respect to `baseline' events defined by requiring \mbox{$p_{T}(W)>200$ GeV}, \mbox{$p_{T}(\mu)>20$ GeV}, \mbox{$p_{T}(\nu)>20$ GeV} and \mbox{$|\eta|(\mu)<5.0$}.
\protect\subref{fig:xsecSig} $WH,W(\mu\nu),H(b\bar{b})$ differential cross-section with respect to $p_{T}(W)$, multiplied by branching ratio and selection efficiency, using the resolved (circles) and substructure (triangles) selections, for $\sqrt{s}=8$ (hollow) and $14$~TeV (solid).}
\end{figure}

The evolution of the signal efficiency for the resolved and substructure methods as a function of $p_{T}(W)$ is 
shown in Fig.~\ref{fig:effSig}. These efficiencies are evaluated after applying the vector boson selection cuts 
described above, and requiring a Higgs boson candidate in the 
invariant-mass window $110<m_{H}<130$~GeV, but before applying the jet veto. 
Efficiencies for events which are {\it uniquely} reconstructed by each approach are shown as dashed lines. 

The resolved method identifies significantly more events than does the substructure approach at lower $p_{T}(W)$ ($200-300$~GeV) 
and approximately 20\% of the events reconstructed in the resolved case are missed by the substructure approach over the full $p_{T}(W)$ range, 
mostly due to a combination of the momentum balance condition of the splitting algorithm, the $B$-hadron-subjet matching requirements, and the mass window condition. 
The two algorithms have very similar performance in the $\sim300-550$~GeV region. 
A marked drop in the efficiency of the resolved method is observed when $p_{T}(W)$ exceeds $600$~GeV, reflecting the 
increasing probability that the $b\bar{b}$ pair be emitted with an angular separation of less than 0.4, and thus
failing to be reconstructed as two \hbox{anti-${k_T}$} $R=0.4$ jets. 

In the $p_{T}(W)>200$ GeV region, events uniquely reconstructed by the substructure approach contribute $\sim$20\% of the total acceptance, a contribution that increases to $\sim$70\% when considering only the $p_{T}(W)>600$ GeV region. 
Considering a luminosity of 150~fb$^{-1}$, this implies an additional $\sim 30$ and $\sim$3 events (in the muon channel alone), respectively. This can be compared to the $\sim$120 and $\sim$1 signal events expected in the resolved case.

The impact of these efficiencies on the accessibility of the signal is demonstrated in Fig.~\ref{fig:xsecSig}, which shows the 
$WH$ differential cross-section with respect to the $W$ transverse momentum, $p_{T}(W)$, multiplied by branching ratio and selection efficiency. 
For comparison, the 8~TeV case is also shown.
For the Standard Model $VH$ process at $\sqrt{s}=14$~TeV, less than 1\% of the signal in the $p_{T}(W) > 200$ GeV region, has $p_{T}(W) > 600$ GeV, where the resolved approach 
begins to fail badly. To measure the high-$p_T$ tail, and for Beyond the Standard Model searches, which expect a significant amount of signal 
at high $p_{T}(W)$, jet substructure techniques are clearly vital. 

\section{Background Estimation}
\label{sec:bkg}

In addition to the signal efficiency, the evolution 
of signal-to-background ratios and significance with $p_{T}(W)$ are important figures-of-merit to conclude on the feasibility of 
the \mbox{$VH,H\rightarrow b\bar{b}$} channel and the usefulness of substructure techniques. 
Bearing in mind the limitations of a particle-level study, 
estimates of identification and reconstruction efficiencies for the main background processes have been made. 

\begin{figure}
\centering
\subfloat[\label{fig:efftt}]{
	\includegraphics[width=0.4\textwidth]{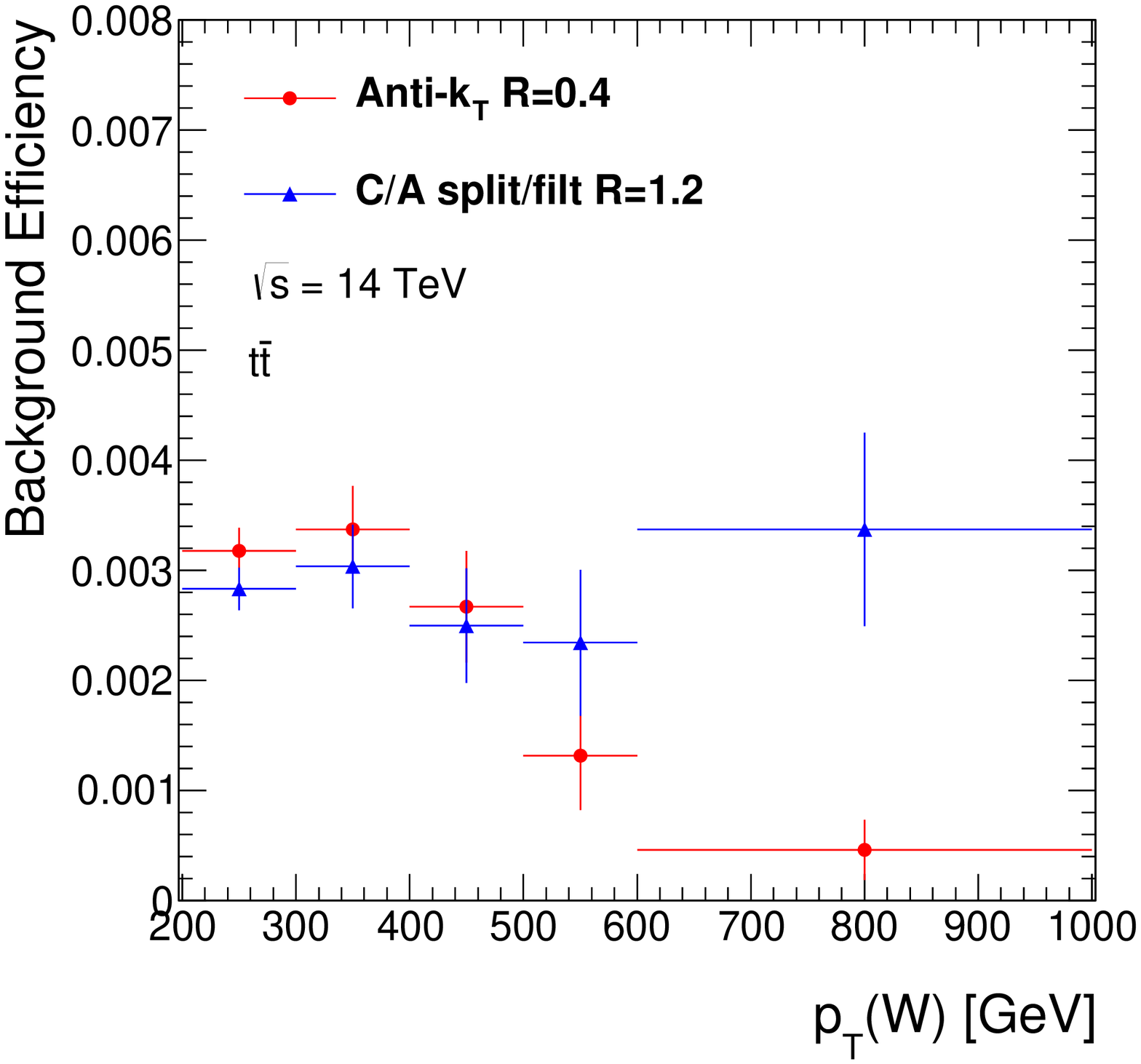}
}
\subfloat[\label{fig:effwbb}]{
	\includegraphics[width=0.4\textwidth]{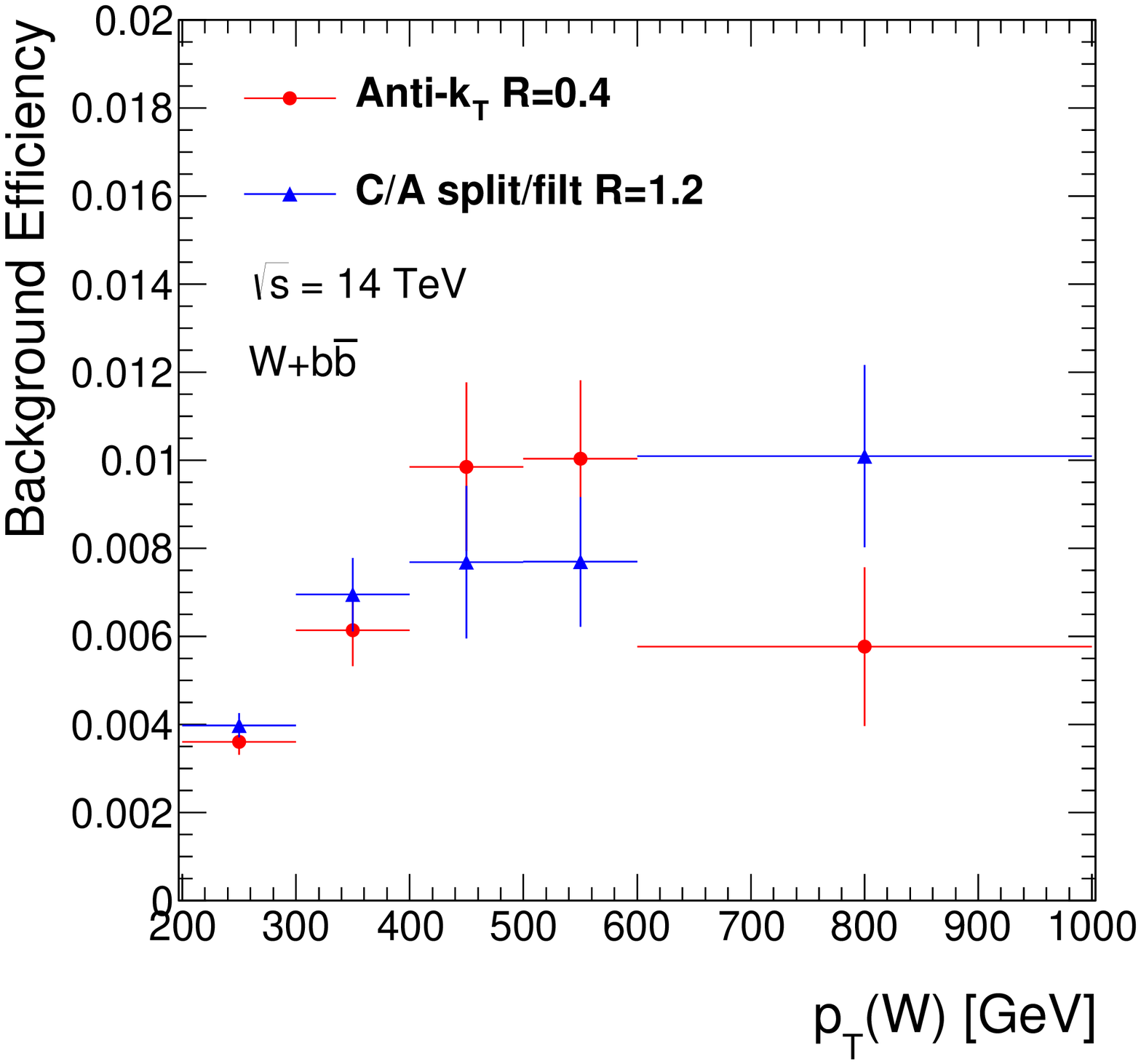}
} \\
\subfloat[\label{fig:effdb}]{
	\includegraphics[width=0.4\textwidth]{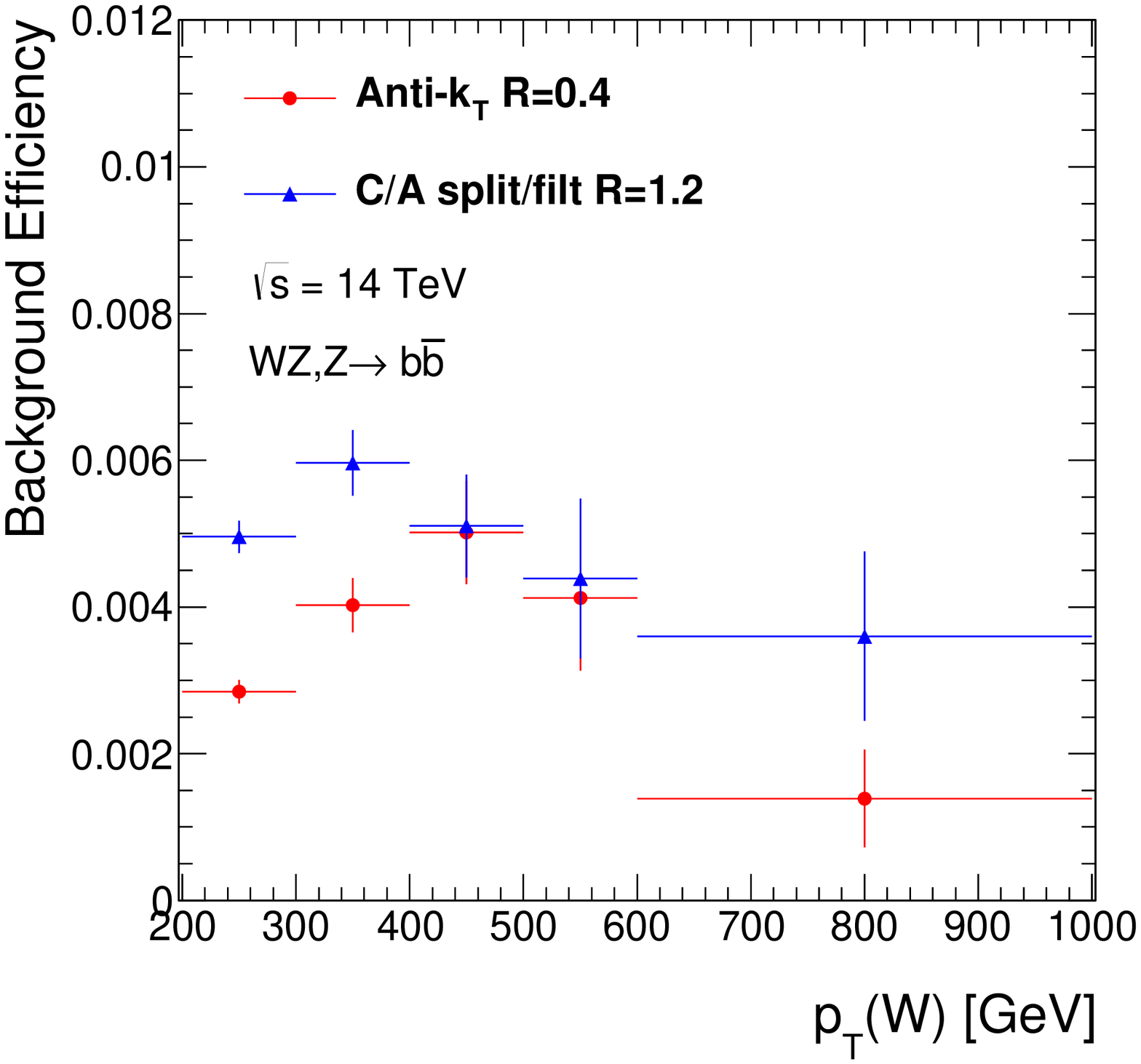}
}
\subfloat[\label{fig:effwt}]{
	\includegraphics[width=0.4\textwidth]{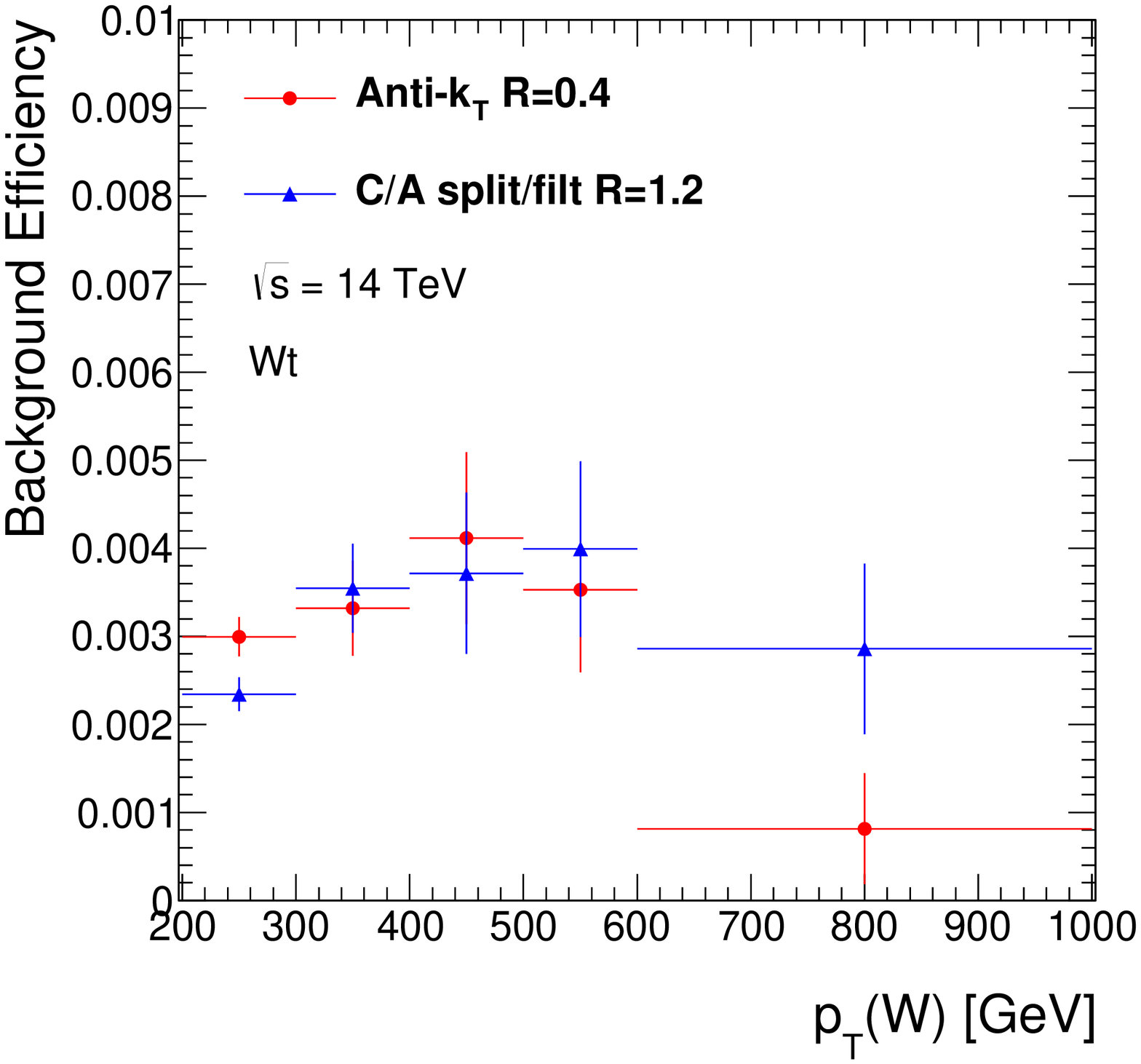}
}	
\caption{\label{fig:effbkg} Background efficiency for the resolved (circles) and substructure (triangles) selections at \mbox{$\sqrt{s}=14$~TeV} for the \protect\subref{fig:efftt} $t\bar{t}$, \protect\subref{fig:effwbb} $W+b\bar{b}$, \protect\subref{fig:effdb} $WZ$ and \protect\subref{fig:effwt} $Wt$ backgrounds. The cuts in Section~\ref{sec:sel} are applied, including the mass window, but excluding the jet veto cuts. The efficiency is evaluated for `baseline' events defined by requiring $p_{T}(W)>200$ GeV, \mbox{$p_{T}(\mu)>20$ GeV}, $p_{T}(\nu)>20$ GeV and \mbox{$|\eta|(\mu)<5.0$}.}
\end{figure}

The background efficiencies for the resolved and substructure methods are shown in Fig.~\ref{fig:effbkg}, as a function of $p_{T}(W)$. 
As with the signal efficiencies in Fig.~\ref{fig:effSig}, they are evaluated after applying the boson selection cuts and mass window but before the jet 
veto. In general the background efficiencies show similar features to the signal, 
with a drop in the resolved efficiency ({\it i.e.} increased rejection) 
around $500-600$~GeV for the resolved method, which is not seen in the substructure method. The exception to this is the $W+b\bar{b}$ background,
where the resolved efficiency does not drop as rapidly. This seems to be due to the fact that wide-angle $b\bar{b}$ pairs produced in the
hard matrix element continue to feed into the boosted kinematic region as $p_{T}$ increases. We also note that below 400~GeV, the $WZ$ background 
is significantly higher in the substructure case, due to $Z \rightarrow b\bar{b}$ decays reconstructed with a mass above 110~GeV.

After the initial event selection, the jet veto rejects roughly 30\% and 40\% of signal events in the Higgs boson mass window with the 
resolved and substructure selection, respectively. It is however extremely effective in reducing the $t\bar{t}$ contamination in the mass 
window rejecting over 90\% of the events in both cases. The efficiency for $W+b\bar{b}$ events is more discrepant between the methods, ranging 
from approximately 30\% to 50\%, with the best rejection achieved by the substructure approach. 

\section{Mass Distributions and Sensitivity}
\label{sec:mass}

\begin{figure}
\centering
\subfloat[\label{fig:mH_inclusive_akt}]{
	\includegraphics[width=0.48\textwidth]{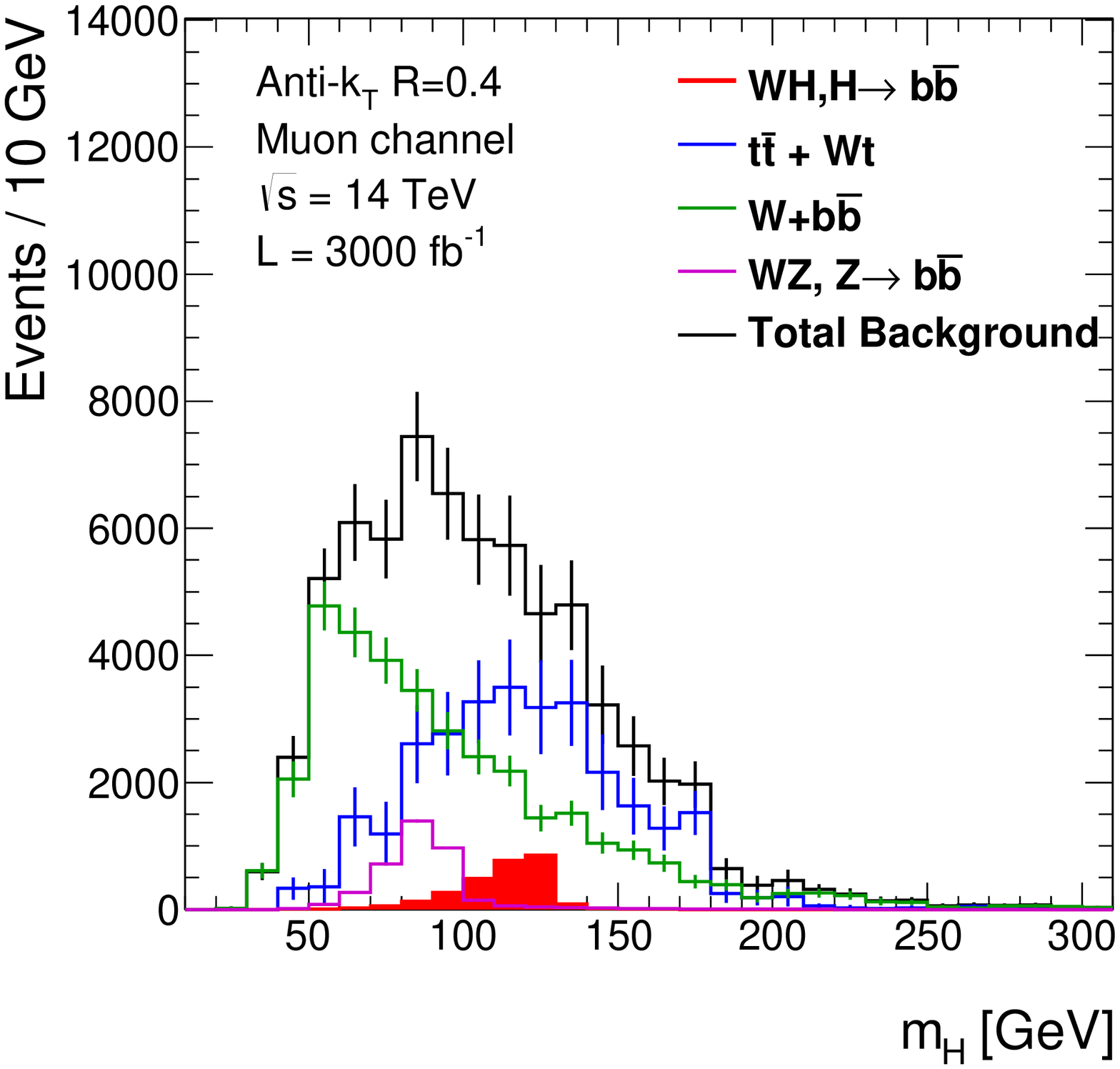}
}
\subfloat[\label{fig:mH_inclusive_JS}]{
	\includegraphics[width=0.48\textwidth]{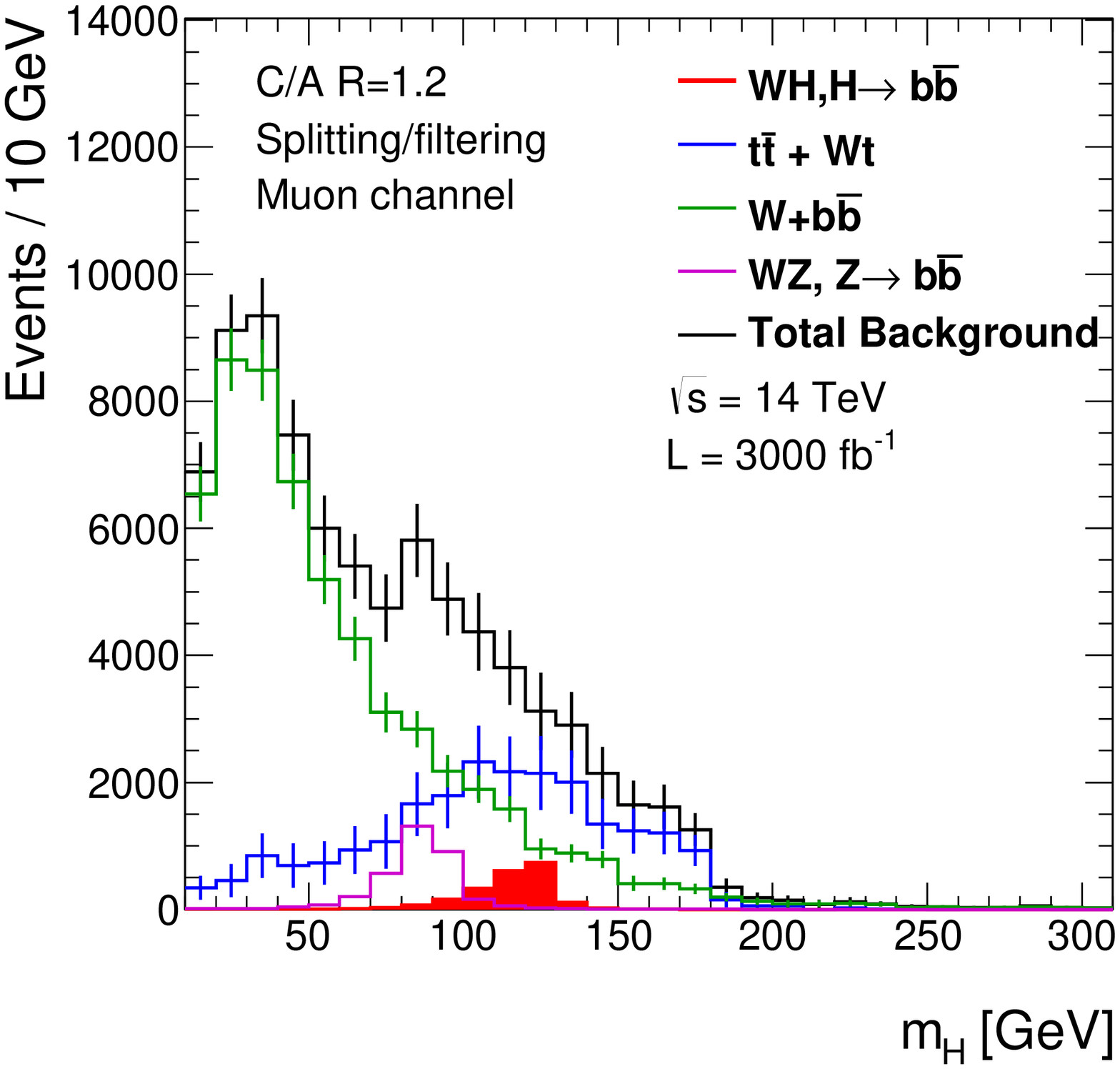}
}
\caption{\label{fig:mH_inclusive} Dijet/Jet Invariant mass after the \protect\subref{fig:mH_inclusive_akt} resolved and \protect\subref{fig:mH_inclusive_JS} substructure selections, including the jet veto, for signal, $t\bar{t}$, $Wt$, $WZ$, and $W+b\bar{b}$ events at $\sqrt{s}=14$ TeV.}
\end{figure}

The invariant mass distributions are shown in Fig.~\ref{fig:mH_inclusive} for both the resolved and substructure approaches for an integrated luminosity of 3000~fb$^{-1}$, with Table~\ref{tab:numEvents} showing the expected number of events in the $m_{H}$ window for each process. The top background has a peak in the same region as the signal, especially in the resolved case. The region of low invariant masses obtained with the substructure reconstruction has a very high purity of $W+b\bar{b}$ events and could in principle be useful as a control region for this background. 

\begin{table}
\centering
\caption{Number of events, for the $W \rightarrow \mu\nu$ channel only, in the Higgs boson mass 
window after the full resolved or substructure selection, including the jet veto, for a luminosity of \mbox{3000~fb$^{-1}$}.}
\label{tab:numEvents}
\begin{tabular}{ccccccc}
\multicolumn{7}{c}{Resolved}\\
\hline\noalign{\smallskip}
$p_{T}(W)$ [GeV] & Signal & $W+b\bar{b}$ & $t\bar{t}$ & $Wt$ & $WZ$ & Total Background \\
\hline\noalign{\smallskip}
$200-400$ & 1405 & 2987 & 5024 & 1165 & 77 & 9253 \\
$400-600$ & 208 & 541 & 361 & 112 & 15 & 1029 \\
$> 600$ & 19 & 89 & 10 & 0 & 1 & 100 \\
\hline\noalign{\smallskip}
$> 200$ & 1632 & 3617 & 5395 & 1277 & 93 & 10382 \\
\hline\noalign{\smallskip}
\multicolumn{7}{c}{Substructure}\\
\hline\noalign{\smallskip}
$p_{T}(W)$ [GeV] & Signal & $W+b\bar{b}$ & $t\bar{t}$ & $Wt$ & $WZ$ & Total Background \\
\hline\noalign{\smallskip}
$200-400$ & 1115 & 2069 & 2718 & 865 & 68 & 5720 \\
$400-600$ & 184 & 278 & 505 & 67 & 9 & 859 \\
$> 600$ & 54 & 184 & 148 & 13 & 3 & 348 \\
\hline\noalign{\smallskip}
$> 200$ & 1353 & 2531 & 3371 & 945 & 80 & 6927 \\
\hline\noalign{\smallskip}
\end{tabular}
\end{table}

Table~\ref{tab:flavorComp} displays the categorisation of events in terms of the flavour composition of the leading and subleading jets: $bb$, $bc$ and $bl$. As expected, the signal is dominated by genuine $b\bar{b}$ events. The $WZ$ and $W+b\bar{b}$ backgrounds are also dominated by $b\bar{b}$, with a few percent contribution from mis-tags. However, most of the $t\bar{t}$ contamination comes from mis-tagged $bc$ events, a component which is even more significant in $Wt$ events\footnote{Consequently, $Wt$ and $t\bar{t}$ produce distributions with similar shapes, and have been merged in all plots.}.

\begin{table}
\centering
\caption{Flavour composition of the events selected by the resolved and substructure selections in the Higgs boson mass window, after the jet veto is applied. The full range \mbox{$p_{T}(W)>200$ GeV} is considered.}
\label{tab:flavorComp} 
\begin{tabular}{cccccc}
\multicolumn{6}{c}{Resolved}\\
\hline\noalign{\smallskip}
Flavour (\%) & Signal & $W+b\bar{b}$ & $t\bar{t}$ & $Wt$ & $WZ$ \\
\hline\noalign{\smallskip}
$bb$ &  99.9 & 93.1 & 32.8 & 7.2 & 94.5 \\
$bc$ & 	0.1 & 4.0 & 55.8 & 78.2 &  3.4 \\
$bl$ &  0.0 & 2.9 & 11.5 & 14.6 & 2.1\\
\hline\noalign{\smallskip}
\multicolumn{6}{c}{Substructure}\\
\hline\noalign{\smallskip}
Flavour (\%) & Signal & $W+b\bar{b}$ & $t\bar{t}$ & $Wt$ & $WZ$ \\
\hline\noalign{\smallskip}
$bb$ & 99.8 & 94.2 & 20.1 & 6.1 & 95.2 \\
$bc$ & 0.2 & 3.2 & 63.7 & 78.9 & 2.9\\
$bl$ & 0.1 & 2.6 & 16.2 & 15.0 & 1.9\\
\hline\noalign{\smallskip}
\end{tabular}
\end{table}

The contribution of $bc$ to the $t\bar{t}$ background also increases as a function of $p_{T}(W)$, making up $\sim85\%$ of the $t\bar{t}$ background in both the resolved and substructure cases for $p_{T}(W) > 400$~GeV. In the resolved case, the $bb$ component becomes negligible in this region, whilst it continues to contribute $\sim5\%$ in the substructure case, with the remaining component due to $bl$.  The $bb$ contribution in the substructure case is composed of a significant fraction of $t\bar{t}$-pairs produced in association with additional heavy flavour jets. This becomes the dominant contribution for $p_{T}(W) > 400$ GeV, where it forms $\sim70\%$ of this background component. Given the large theoretical uncertainties on such production, this could add an additional level of difficulty in probing this region of phase space. In the resolved case there is a negligible fraction of $t\bar{t}$-pairs produced in association with additional heavy flavour jets in all $p_{T}(W)$ regions.

Improvements in $b$-tagging techniques, in both improving their level of charm-quark rejection and increasing the acceptance to identify additional $b$-jets in the events, are vital to reduce the $t\bar{t}$ contribution in the mass window of the Higgs boson.

\begin{figure}
\centering
\subfloat[\label{fig:mH_bin1_akt}]{
	\includegraphics[width=0.31\textwidth]{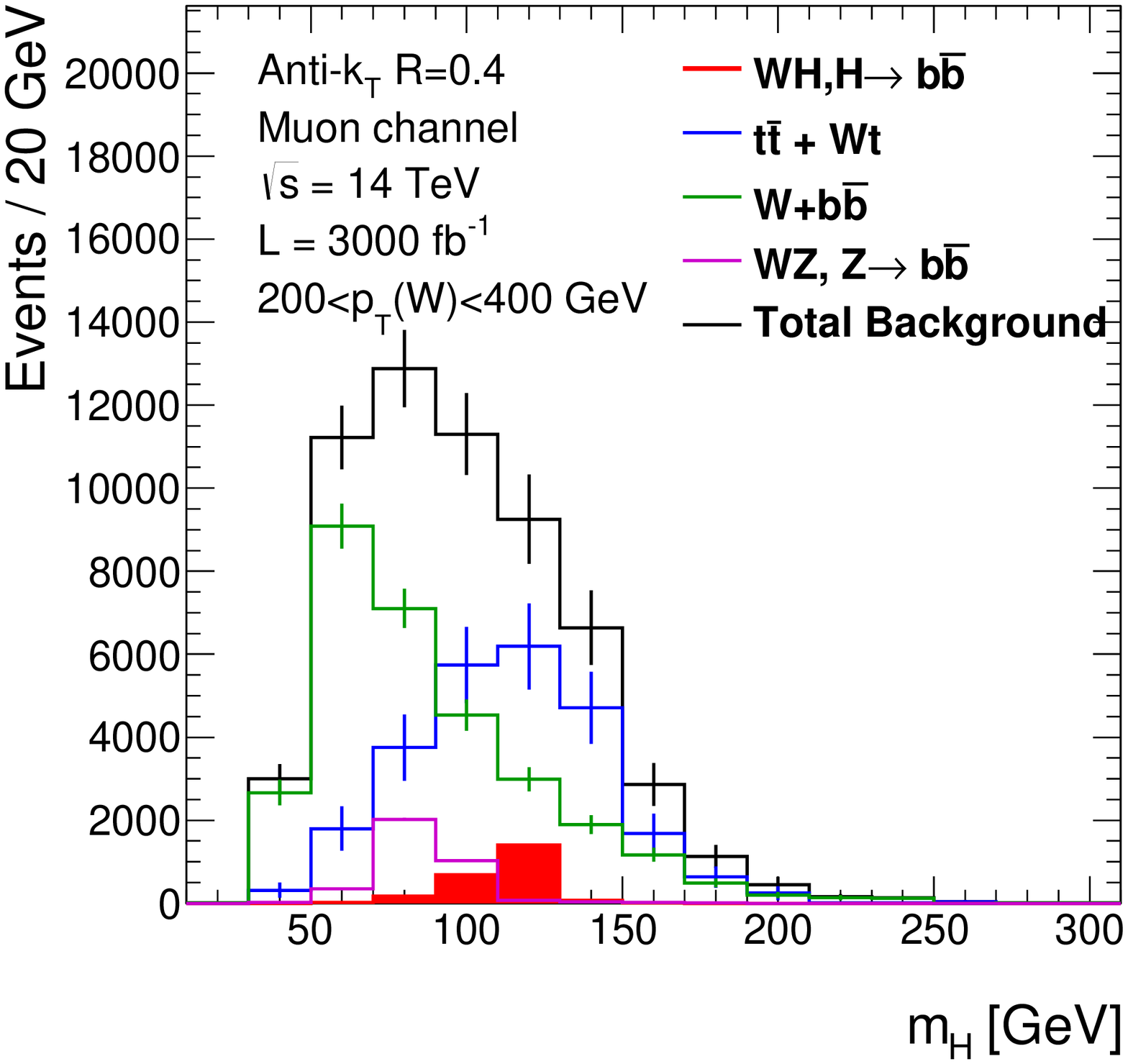}
}
\subfloat[\label{fig:mH_bin2_akt}]{
	\includegraphics[width=0.31\textwidth]{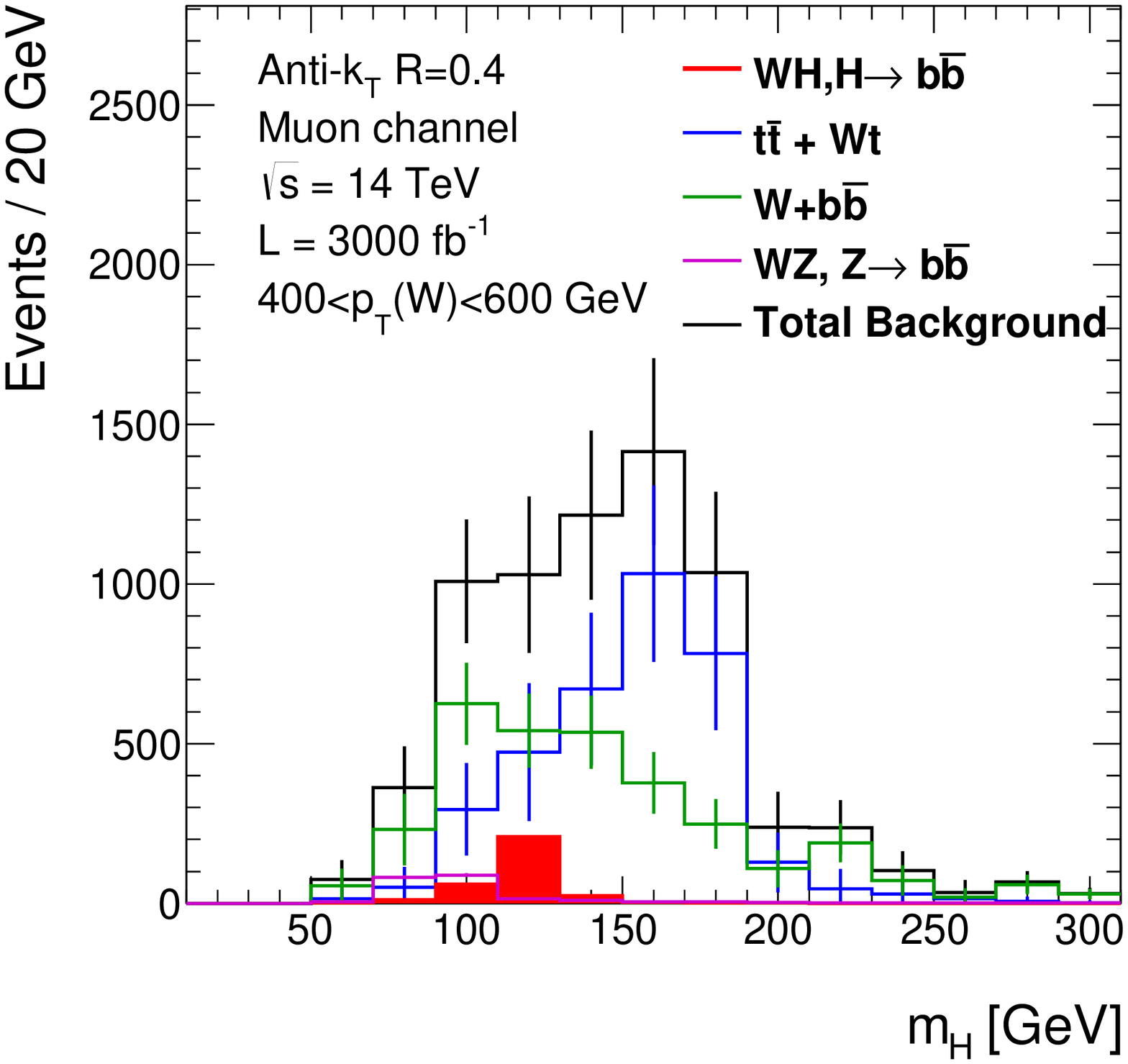}
}
\subfloat[\label{fig:mH_bin3_akt}]{
	\includegraphics[width=0.31\textwidth]{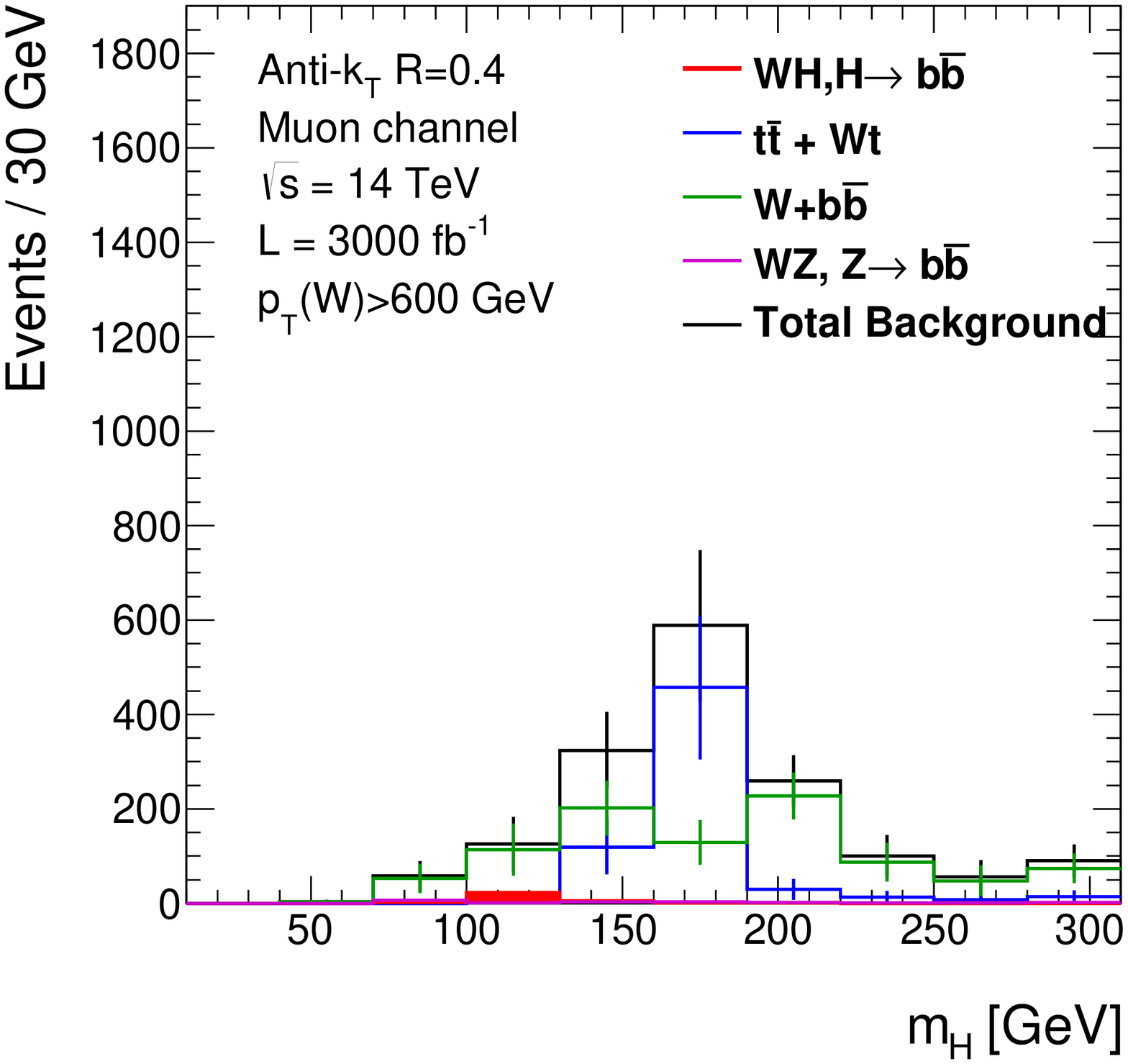}
}\\
\subfloat[\label{fig:mH_bin1_JS}]{
	\includegraphics[width=0.31\textwidth]{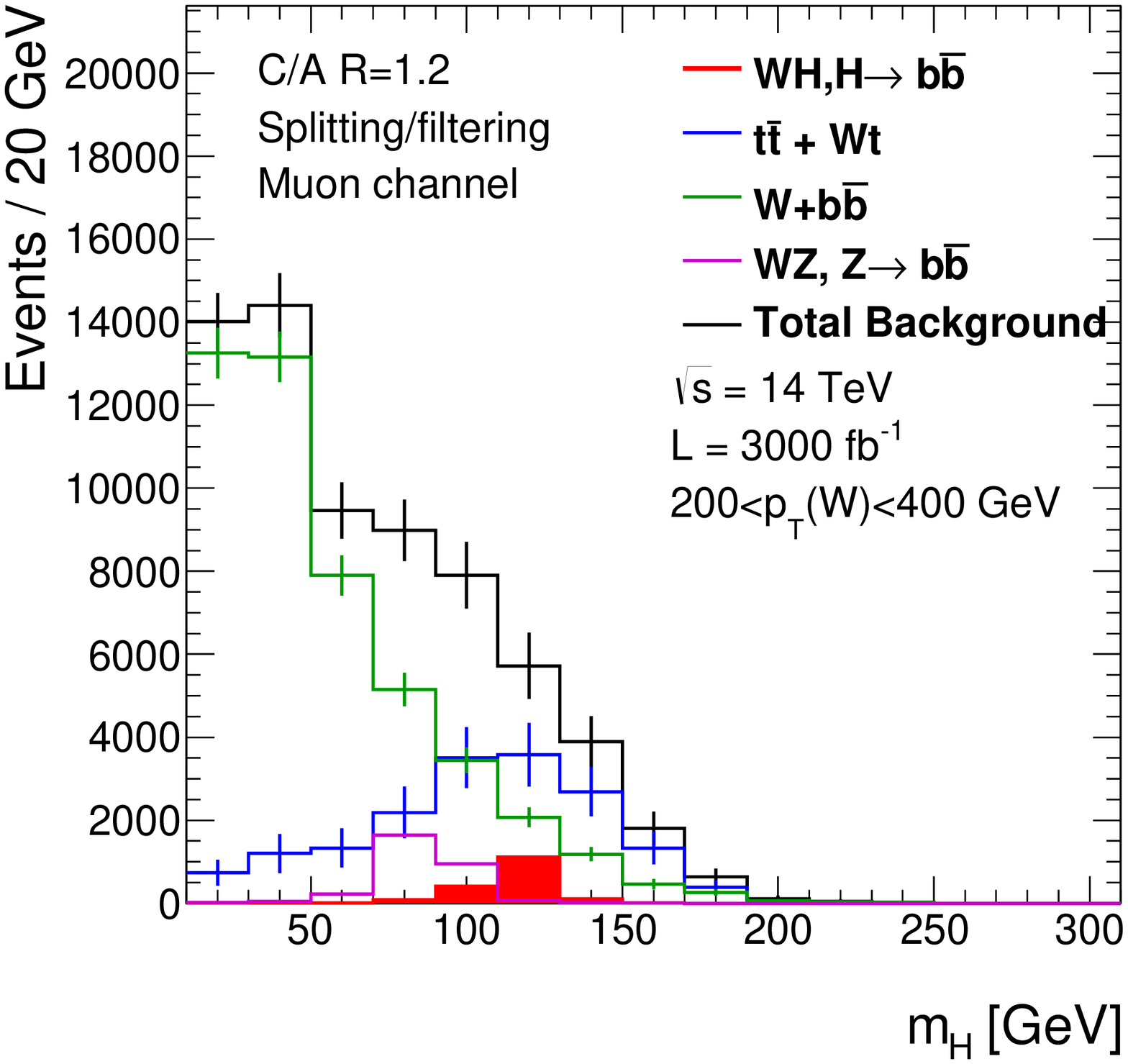}
}
\subfloat[\label{fig:mH_bin2_JS}]{
	\includegraphics[width=0.31\textwidth]{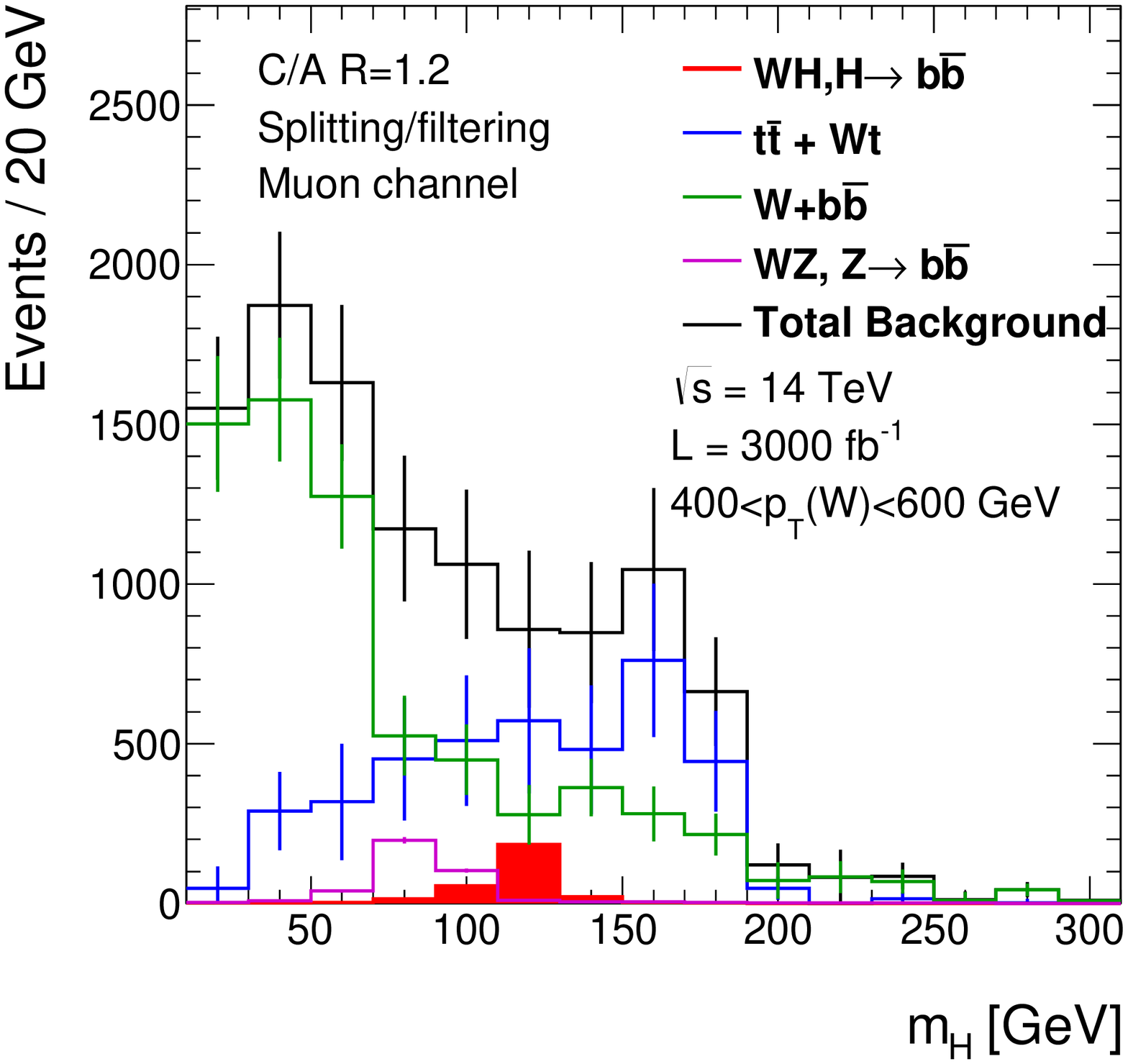}
}
\subfloat[\label{fig:mH_bin3_JS}]{
	\includegraphics[width=0.31\textwidth]{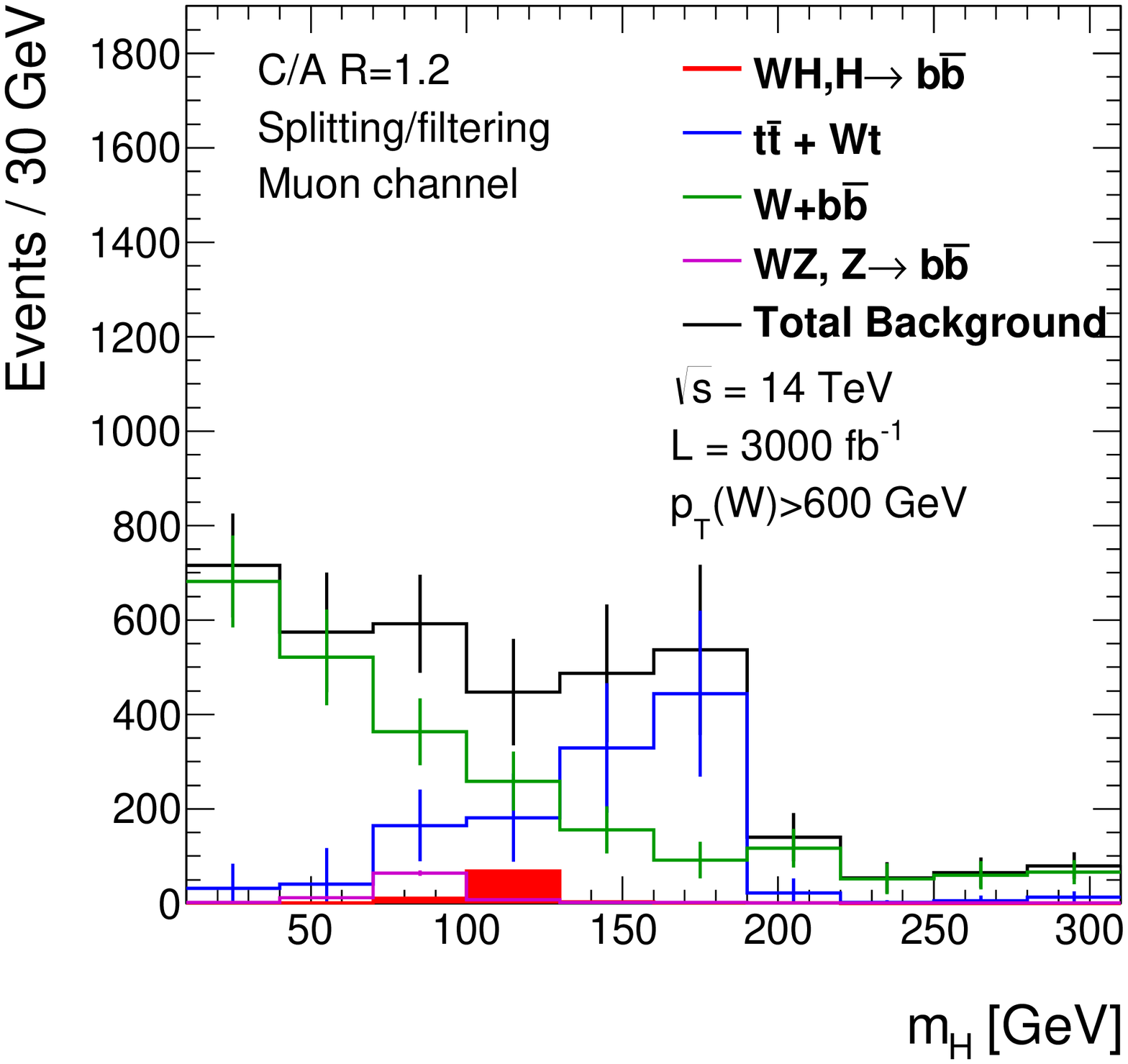}
}
\caption{\label{fig:mH_WpT} Dijet/Jet invariant mass for the \protect\subref{fig:mH_bin1_akt}-\protect\subref{fig:mH_bin3_akt} resolved and \protect\subref{fig:mH_bin1_JS}-\protect\subref{fig:mH_bin3_JS} substructure selection including the jet veto, for signal, $t\bar{t}$, $Wt$, $W+b\bar{b}$ and $WZ$ events at $\sqrt{s}=14$ TeV, for $p_{T}(W)$ \protect\subref{fig:mH_bin1_akt}+\protect\subref{fig:mH_bin1_JS} $200-400$, \protect\subref{fig:mH_bin2_akt}+\protect\subref{fig:mH_bin2_JS} $400-600$ and \protect\subref{fig:mH_bin3_akt}+\protect\subref{fig:mH_bin3_JS} $>600$ GeV.}
\end{figure}

All cross-sections fall rapidly with increasing $p_{T}(W)$, and the evolution of rates and shapes can be seen in Fig.~\ref{fig:mH_WpT} for the resolved and substructure cases. Despite the limited statistics, it is observed that in the resolved analysis, the shapes of the $W+b\bar{b}$ and $t\bar{t}(bb)$ background processes are kinematic in origin, and heavily dependent on the boost of the system. 

An estimation of the signal sensitivity for both the resolved and substructure approaches is made, assuming integrated luminosities of 150 and 3000~fb$^{-1}$, corresponding to expectations for Run 2 of the LHC and for the eventual goal of a high luminosity upgrade. As well as the muon channel studied above, signal and background events originating from the electron decay channel are also taken into account, assuming the same acceptance.  

The signal-to-background ratios are shown in bins of $p_{T}(W)$ in Table~\ref{tab:results}, calculated in the Higgs boson mass window. The substructure method achieves a higher $S/B$ in the \mbox{$200< p_{T}(W)<400$~GeV} range, and the values for higher boosts are compatible between the two methods, within the statistical uncertainties. Given the significant drop in signal efficiency obtained with the resolved approach for values of $p_{T}(W)$ greater than \mbox{600 GeV}, a decrease in $S/B$ might have been expected. However, this drop is accompanied by a similar decrease in the background efficiency. 

\begin{table}
\centering
\caption{Signal-to-background ratio and signal significances in the full boosted range and in each $p_{T}(W)$ bin. The figures of merit are calculated considering all events selected by the resolved and substructure selections, and also events that were uniquely selected by the latter, after the jet veto is applied. The acceptance from the electron channel is taken into account. }
\label{tab:results} 
\begin{tabular}{cccc}
\multicolumn{4}{c}{$S/B$(\%)} \\
\hline\noalign{\smallskip}
\multirow{2}{*}{$p_{T}(W)$ [GeV]} & \multirow{2}{*}{Resolved} & \multirow{2}{*}{Substructure} & Unique \\
 &  &  & Substructure \\
\hline\noalign{\smallskip}
$200-400$ & 15.2 & 19.5 & 8.7 \\
$400-600$ & 20.3 & 21.5 & 6.0 \\
$>600$ & 19.2 & 15.6 & 13.9 \\
\hline\noalign{\smallskip}
$>200$ & 16.0 & 19.9 & 9.1 \\
\hline\noalign{\smallskip}
\multicolumn{4}{c}{$S/\sqrt{B}$, $L=3000(150)$~fb$^{-1}$} \\
\hline\noalign{\smallskip}
\multirow{2}{*}{$p_{T}(W)$ [GeV]} & \multirow{2}{*}{Resolved} & \multirow{2}{*}{Substructure} & Unique \\
 &  &  & Substructure \\
\hline\noalign{\smallskip}
$200-400$ & 20.6 (4.6) & 20.8 (4.7) & 4.7 (1.1) \\
$400-600$ & 9.2 (2.1) & 8.9 (2.0) & 1.6 (0.4) \\
$>600$ & 2.7 (0.6) & 4.1 (0.9) & 3.4 (0.8) \\
\hline\noalign{\smallskip}
$>200$ & 22.7 (5.1) & 23.0 (5.1) & 5.9 (1.3) \\
\hline\noalign{\smallskip}
\end{tabular}
\end{table}

The $S/\sqrt{B}$ is calculated in bins of $p_{T}(W)$, as shown in Table~\ref{tab:results}. It is observed that the most significant event region corresponds to the range \mbox{$200< p_{T}(W)<400$}~GeV, where the resolved approach continues to perform well, and that higher boosts do not help in achieving a higher signal significance. This observation suggests that the great advantage in boosting the $VH$ system consists in reducing the combinatorial background and the large $t\bar{t}$ contribution, achieved with transverse momenta on the order of the Higgs boson mass. Higher $p_{T}$ values are not beneficial to the signal significance due to the extremely small signal cross-section.  

\begin{figure}
\centering
\subfloat[\label{fig:result_bin1}]{
	\includegraphics[width=0.31\textwidth]{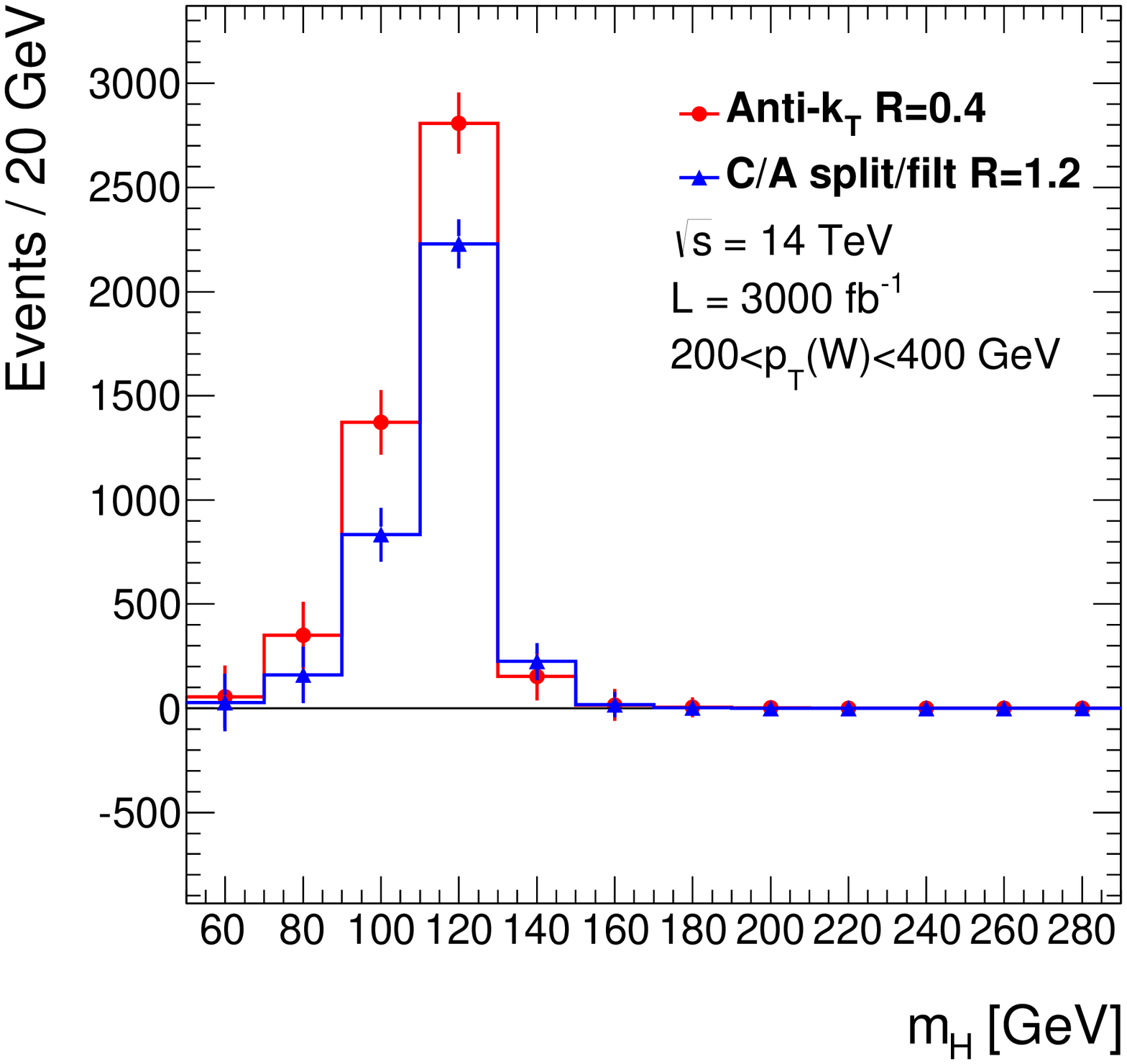}
}
\subfloat[\label{fig:result_bin2}]{
	\includegraphics[width=0.31\textwidth]{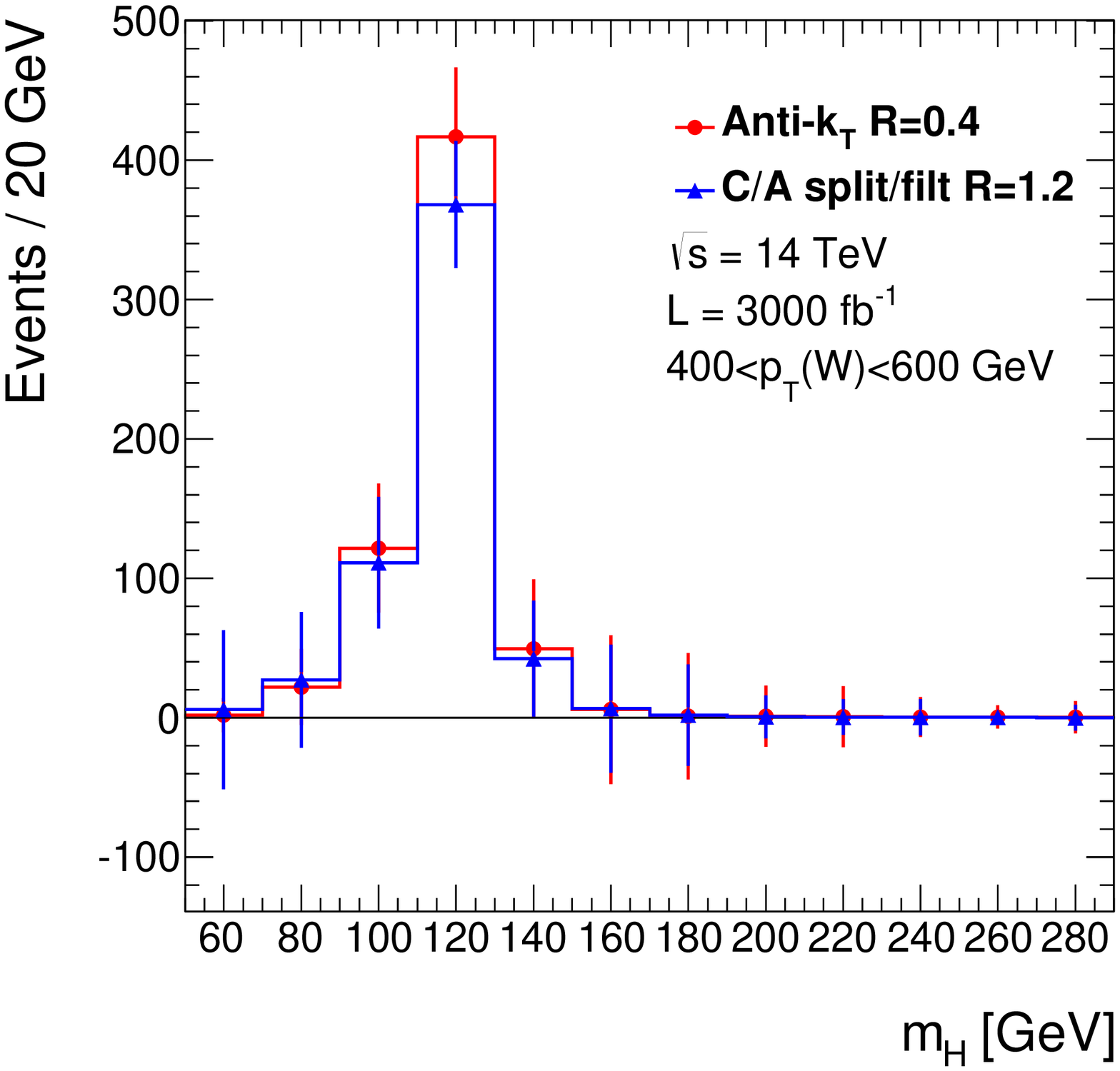}
}
\subfloat[\label{fig:result_bin3}]{
	\includegraphics[width=0.31\textwidth]{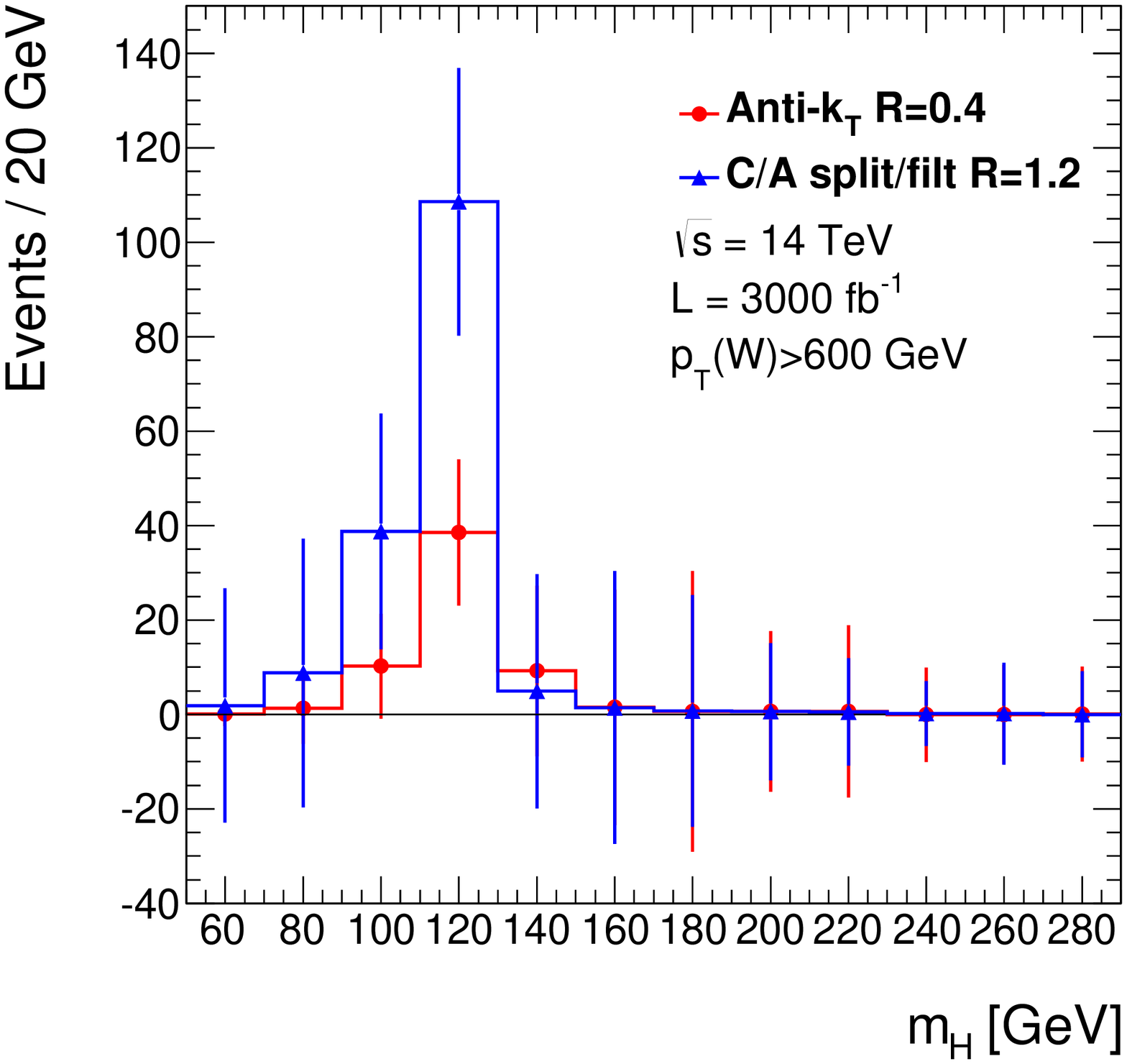}
}
\caption{\label{fig:result} Dijet/Jet invariant mass for the resolved (circles) and substructure (triangles) selections, including the jet veto, after subtraction of all backgrounds, for $p_{T}(W)$ \protect\subref{fig:result_bin1} $200-400$, \protect\subref{fig:result_bin2} $400-600$ and \protect\subref{fig:result_bin3} $>600$ GeV.} 
\end{figure}

The two analyses achieve similar significances in the range \mbox{$p_{T}(W)<600$ GeV}, while the substructure approach outperforms in the highest bin, increasing the significance by approximately 50\%. A combination of the events reconstructed by the resolved approach with those uniquely reconstructed by the substructure approach has the potential to increase the significance of the highest $p_{T}(W)$ region by approximately $\sim$60\%. A Run 2 measurement targeting the full boosted regime can already achieve a statistical significance of 5$\sigma$, a result that could be improved by a few percent by combining both the resolved and substructure methods. Figure~\ref{fig:result} shows the expected background-subtracted signal mass-peak for a luminosity on 3000 fb$^{-1}$, with error bars illustrating the anticipated statistical uncertainty. The information from both approaches could also be combined in more sophisticated ways, such as a multivariate technique, to take advantage of the complementary information such techniques can provide to better reject and control the main background processes. 

This study considers only the $WH$ channel, without systematic uncertainties. The addition of the $ZH$,$H\rightarrow b\bar{b}$ channels, for the cases of $Z$ decaying to either leptons or neutrinos, will significantly increase the statistical sensitivity. Additionally, further optimisations of the event selection can also be expected to further improve the sensitivity. The inclusion of systematic uncertainties will degrade the sensitivity, although given the large datasets available, it should be possible to control such uncertainties to a higher degree than was the case in Run 1 of the LHC. The conclusions reached on the relative applicability of the resolved and jet substructure approaches should not be strongly dependent on either of these consideration though. There is however an indication from these studies, that as the substructure approach gives a higher $S/B$ in the most sensitive region, as well as a rather pure $W+b\bar{b}$ control region which could be used to constrain that background, it could have improved sensitivity relative to the resolved case once systematic uncertainties are included (assuming the two approaches have similar sensitivity to the main nuisance parameters in a profile likelihood fit and the experimental uncertainties related to the jets are comparable).

A comparison between this study and previous work~\cite{ref:boostedhiggsprl} indicates that the substructure results for the $WH$ here are consistent, apart from the fact that the $Wt$ background and $bc$ contamination are better estimated here (as was also done by ATLAS using a full detector simulation~\cite{atlas:boostedhiggs}). The principle new factors which make the benefits of using jet substructure less dramatic are the 125~GeV mass of the Higgs boson and the excellent performance of the \hbox{anti-${k_T}$} algorithm over the $200-400$~GeV range. 

\section{Conclusions}
\label{sec:conclusions}

An updated feasibility study of a $WH,H\rightarrow b\bar{b}$ search at a $pp$ collider has been performed exploring the centre-of-mass energy of $\sqrt{s}=14$~TeV in the boosted regime, using both a resolved dijet and jet substructure selection to reconstruct the Higgs boson candidate. The most sensitive region is found to be $p_{T}(W)=200-400$~GeV, with higher $p_{T}(W)$  regions not improving the statistical sensitivity. In this region, both jet selections perform well. However, for $p_{T}(W) > 600$~GeV, the substructure analysis is essential to retain signal efficiency and sensitivity; this region is of interest for Standard Model measurements at high luminosities and for searches Beyond the Standard Model. Combining both approaches over the full range could also be expected to bring additional benefits. As expected, $b$-tagging is a central issue, especially given that the $t\bar{t}$ contamination comes mainly from mis-tagged charm jets. 

In summary, the measurement of $H \rightarrow b\bar{b}$ decays in the $VH$ production channel remains challenging, but possible, 
in 14~TeV running of the LHC. Either a resolved dijet or jet substructure selection work equally well for the most sensitive regions, but to obtain maximum sensitivity and to probe the $p_T$ dependence, 
both approaches are important.

\section*{Acknowledgments}

The authors would like to thank Andy Buckley and Rikkert Frederix for useful discussions. This work was supported in part by STFC and the Royal Society (UK), the FCT (Portugal) and the Research Executive Agency (REA) of the European Union under the Grant Agreement PITN-GA2012-316704 (``HiggsTools'').

\bibliographystyle{h-physrev4}
\bibliography{hbb_14TeV}

\end{document}